# FUTURE OF ARTIFICIAL INTELLIGENCE IN AGILE SOFTWARE DEVELOPMENT

[1]**Mariyam Mahboob** [2]**Mohammed Rayyan Uddin Ahmed**, [3]**Zoiba Zia**, [4]**Mariam Shakeel Ali**, [5]**Ayman Khaleel Ahmed**

[1]Student [IT Department], Muffakham Jah College of Engineering and Technology, Hyderabad, India
[2]Student [CSE Department], Muffakham Jah College of Engineering and Technology, Hyderabad, India
[3,5]Student [AI-DS Department], Muffakham Jah College of Engineering and Technology, Hyderabad, India
[4]Student [IT Department], Muffakham Jah College of Engineering and Technology, Hyderabad, India

Email:[1]mariyam.mn13@gmail.com [2]rayyan9290@gmail.com, [3] zoiba.zia27@gmail.com, [4] maryum2125@gmail.com, [5]khaleelayman47@gmail.com

Contact:[1]+91 8106039724 [2]+91 9059598390, [3]+91 9581554276, [4]+91 7032115892, [5]+91 8106290579

**Abstract:** The advent of Artificial intelligence has promising advantages that can be utilized to transform the landscape of software project development. The Software process framework consists of activities that constantly require routine human interaction, leading to the possibility of errors and uncertainties. AI can assist software development managers, software testers, and other team members by leveraging LLMs, GenAI models, and AI agents to perform routine tasks, risk analysis and prediction, strategy recommendations, and support decision-making. AI has the potential to increase efficiency and reduce the risks encountered by the project management team while increasing the project success rates. Additionally, it can also break down complex notions and development processes for stakeholders to make informed decisions. In this paper, we propose an approach in which AI tools and technologies can be utilized to bestow maximum assistance for agile software projects, which have become increasingly favored in the industry in recent years.

## I. INTRODUCTION

AI has certainly transformed our lives in ways that we cannot comprehend. Due to reduced computational costs and breakthroughs in algorithms, AI has excelled in domains like decision-making, pattern recognition, and natural language processing. Consequently, AI has become increasingly integrated into software engineering, effectively minimizing human errors [1]. Project management is a discipline that seeks to optimize a team's efforts and resources in the development of a project through various models that provide a framework for team management, decision-making, risk mitigation, and customer satisfaction. One popular software project management model that has gained acceptance in the industry is the agile development model. Agile development methodology relies heavily on the concept of a product backlog, which serves as a repository of user requirements that expand as demands evolve. A backlog is further divided into smaller, actionable tasks that are scheduled by the development team, leading to project delivery in incremental milestones. This approach effectively mitigates risks by segmenting the development costs into multiple stages or sprints. The agile process encompasses various iterations or sprints, starting from gathering user stories and feedback to ideation, planning, and execution.

Although numerous tools that assist the agile framework exist, for example, JIRA which aids in data collection and task management, decision-making still predominantly relies on human intervention, a gap that artificial intelligence (AI) aims to bridge. By leveraging AI tools such as Large Language Models (LLMs), autonomous agents, and AI algorithms, systems can simulate risk management, project trajectory planning and provide estimated pros and cons while continuously learning from decisions made throughout the development stages

## II. HISTORY

Over the last several decades, the software industry has undergone many developments and transformations in numerous software development methodologies. Each method possessed eccentric qualities that distinguished it from other methods. One of the methods was the Traditional Software Development Method (TDSM), which was largely composed of waterfall and spiral methods that depended on well-organized planning, process, documentation, and comprehensive design. The TSDMs are still widely used in industry because of their straightforward, organized, and structured nature and their predictability, stability, and high assurance [2]. Though many TSDMs have been developed since the waterfall model to provide significant productivity improvements, none are free from major problems including blown budgets, missed schedules, and flawed products. They have failed to provide dramatic improvements in productivity, in reliability, and simplicity [3].

## III. CHALLENGES IN THE AGILE SOFTWARE DEVELOPMENT PROCESS

### A. Agile Software Development Methods

To address the limitations of Traditional Software Development Models (TSDMs), numerous Agile Software Development Methods like Scrum, Extreme Programming (XP), and Lean Software Development (LSD) have evolved which focus on iterative and incremental development, customer collaboration, and frequent delivery[4] through a light and fast development life cycle. Although many benefits of agile approaches including shorter development cycles, higher customer satisfaction, lower bug rate, and quicker adaptation to changing business requirements have been reported [2], XP has emerged as a solution to issues that existed due to the long development cycles of traditional models. Despite constant changes in requirements in small and medium-sized teams, it's proven successful in software development. The main characteristics of XP include customer participation, coordination, communication, quick feedback, precise documentation, and pair programming. Lean Software Development (LSD) is an iterative methodology that focuses on reducing waste and optimizing the entire process to achieve the maximum possible gain[8]. It focuses on quick and efficient feedback between customers and software developers to attain better productivity, workflow, and development. It does not adhere to firm guidelines and therefore is considered to be one of the most flexible Agile Methods.

### B. Issues and Challenges

XP is arguably one of the best methods provided that the project has to be completed in a shorter time frame. It prioritizes a fast-paced working environment for prototyping, unlike other methods that consume more time. The XP programmers emphasize and prioritize software coding tasks rather than documentation tasks. Since numerous changes cannot be documented systematically, there is a high possibility of unexpected failures that cannot be tracked. Lack of precise documentation leads to the recurrence of bugs and errors that were resolved earlier. Furthermore, In extreme programming, meetings are held frequently with the customers/investors regarding the progression made on tasks daily. This serves to be an issue as it costs additional time while meeting the programmers face-to-face can result in exhaustion when the customer has a fairly distant location.

The main idea of coding (and one of the most XP aspects) is to implement pair programming. It is recommended by the XP that a development team of two members share one computer and implement side-by-side software. One developer writes the code, while the other developer challenges, supports, and observes the selected method to obtain better results [5]. This extensive pair programming increases the expenses as it is not required for small teams with a limited budget. Overall, the Extreme Programming method requires lots of effort, persistence, and patience as tight deadlines are required to be accomplished every day.

On the other hand, Lean Software Development focuses on the project management aspects of a project and specifies no technical practices; it integrates well with other agile methodologies, such as XP, that focus more on the technical aspects of software development [7]. A major challenge faced here is that it involves comprehensive and precise information at every stage that results in documenting and recording every stage diligently which is hard to keep track of. Secondly, the prime objective is to eliminate waste and optimize the entire process to achieve the maximum possible gain [8]. It doesn't encourage the idea of heavy documentation and usage of diagrams so this process of elimination takes up a huge amount of time and energy resulting in slow development. Lean Software Development encourages flexibility of ideas from customers, but this leads to issues and challenges that create complexity and loss of originality. Lastly, Lean Software Development heavily relies on the team members. If the team is found inconsistent with their skills, this inability leads to inefficiency and affects the overall productivity. The conditions are thus similar to the Traditional Software Development Methods.

## IV. METHODOLOGY

### A. Enhancing Agile Development Through Generative Artificial Intelligence

The above obstacles in Agile Software Development present serious issues in the efficiency and product development quality at the cost of more labor and time. These serious issues can be significantly reduced by integrating modern artificial intelligence at various stages of the agile software development process, such as Extreme Programming (XP) and Lean Software Development (LSD). Modern Artificial intelligence, which consists of Generative Artificial Intelligence, AI agents, AI algorithms, and LLM-based applications, can streamline the agile software development process by offering innovative solutions to enhance efficiency, collaboration, and adaptability. Generative AI, through its ability to analyze patterns and generate deterministic ideas, can assist teams in decision-making, and ideation, and can overcome hurdles caused by human errors while also nurturing a culture of continuous, uninterrupted innovation [9].

However, that is not to say, such tools haven't existed in the past. Tools similar to the above have existed, but they're limited in their abilities to assist the development and management team. The existing agile tools such as JIRA, Assembla, and Axosoft help with project management that is user stories, product backlogs, sprints, and sprint backlogs [10]. The existing tools do not go beyond the scope of management to make agile software project development and management a seamlessly integrated process with minimal erroneous cases.

The emergence of numerous generative software platforms enables the conversion of textual commands to programming code. Developmental tools such as Visual Studio Code, JetBrains IDE, and other IDEs can be expanded in their capabilities to assist the programmer with the integration of GitHub Copilot. GitHub Copilot is an AI assistant built on top of OpenAI Codex, a generative AI system developed by OpenAI.

The recent release of GPT-4o has propelled companies like Microsoft to fuse AI with operating systems to make the development of new applications and software much easier [11]. IBM's Watson, which is renowned for its natural language processing and cognitive computing, has also been implemented for many obstacles faced in the software engineering process. Watson leverages AI to analyze unstructured data which includes but is not limited to documentation, forums, and code repositories. It also supports developers by providing insights and helping them navigate complex codebases [12].

Likewise, the integration of generative AI models can mitigate developmental challenges and project management challenges in the agile software development process. Some of the key features of XP are pair programming, test-first, coding standards, continuous integration, metaphor, and refactoring. Generative AI models like GitHub Copilot can be integrated to assist the programmer as well as test the code as soon as it's written to check if there are no errors, as a result reducing the labor required. The integration of update features when the team reaches certain milestones the stakeholders can ensure customer involvement throughout the process. Furthermore, the AI system can also provide simple explanations, documentation, and visual instructions to the customers, project managers, and other programmers which makes the process of transferring from one stage of the development process to another process. This can be done through a LLM-based AI called Sora AI. Depending on the workflow of the ongoing project, the AI can also estimate the time required to complete the project based on the requirements input into the model generating the test code. We can implement Generative AI models likewise to other agile process models like LSD to reduce the workload on programmers, make the process time-efficient, minimizing the number of iterations needed all while keeping the stakeholders in the loop. This eliminates waste, optimizes the project to deliver the best, and makes it easier for team members to work with code while making it easier to scale without inciting conflict among the team members. The project is well-documented and tested at each stage. Thus, as a result, making the agile software development and management process emphasizes the principles that it is laid upon while minimizing the cons involved in the process.

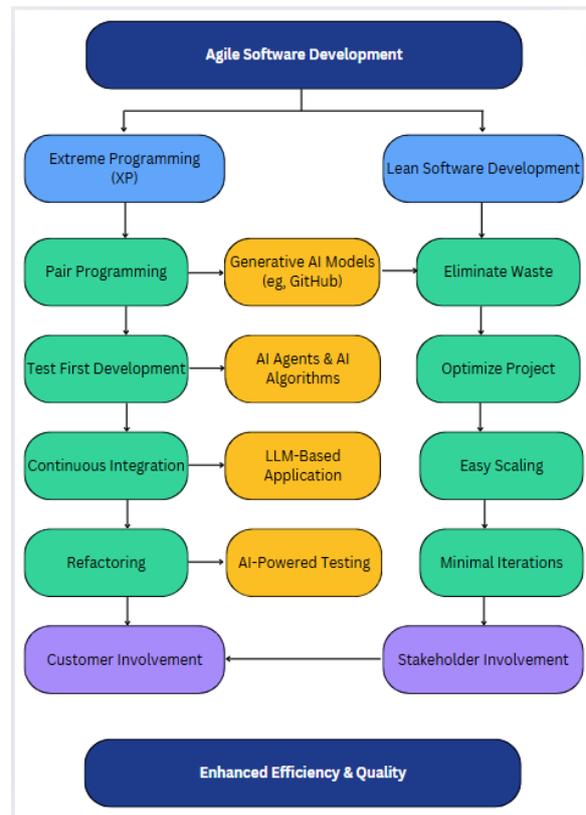

**Figure 1: Integration of AI into Agile Software Development**

### B. Integration of Artificial Intelligence Algorithms in Agile Software Development Lifecycle

The agile development life cycle consists of five major phases: Initiation, planning, execution, monitoring & control, and closing [13]. This section proposes to utilize AI algorithms to automate and enhance the Agile SDLC. As the adoption of Generalized AI becomes more prevalent throughout various industries, as highlighted by [14], problems in the domain of software development can be solved utilizing existing intelligent algorithms. This can be achieved by translating the agile process to the space of machine learning effectively connecting problems with existing solutions[15]. The first phase initiation can utilize supervised machine learning to predict the project success rate based on resources, efforts, cost, and other critical factors. The second phase i.e. planning can leverage supervised machine learning to assign tasks to team members based on the previous performance data. The third and fourth stages closely relate to tasks that can leverage generative AI capabilities as discussed in the above and below sections in great detail. Tasks like report generation, data analysis, and automation would come under these phases.

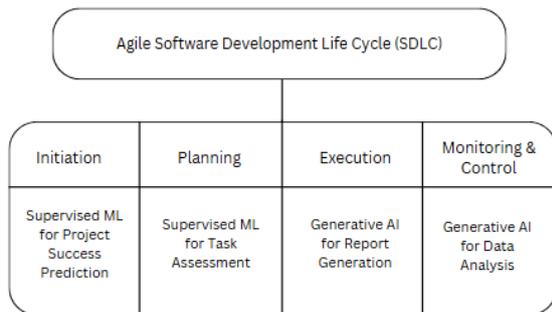

**Figure 2: Integration of AI Algorithms in Agile Software Development Lifecycle**

### C. The Role of Artificial Intelligence in Extreme Programming and Lean Software Development

In the agile methodology, the two of the most significant frameworks are Extreme Programming (XP) and Lean Software Development (LSD). This section explores the various use cases of AI algorithms in the frameworks mentioned above.

I. Use Cases of AI in Extreme Programming (XP):
a. Automated Testing and Debugging:
The usage of AI can help developers evade inefficient code practices by utilizing patterns from historic data[16] and help generate test cases.
b. Risk Assessment:
Predictive AI algorithms can be used to identify potential risks and issues before they escalate, allowing teams to implement mitigation strategies preemptively [17].
c. Continuous Integration and Deployment:
AI can be used in the optimization of product deployment leading to minimal deployment-related issues[18].

II. Use cases of AI in Lean Software Development (LSD):
a. Process Optimization:
AI can provide valuable insights and iterative advice to streamline workflow and prevent redundant effort. By the use of specific Machine Learning models, we can predict when systems or tools are likely to fail and halt the development process. Using such a method allows us to take appropriate measures to prevent any system failure that could halt the development process [19].
b. Performance Monitoring:
AI can continuously analyze and track system performance and user interactions. By monitoring this data, AI can provide necessary improvements that can help the development teams deliver the best quality software consistently.
c. Quality Assurance:
Through the use of AI-powered tools, the quality of the code can be managed. Such tools can continuously analyze the code and assess it for possible vulnerabilities, hence reducing the risk of failure.

### V. CONCLUSION

Our research indicates that the current software development life cycle for agile processes can be further enhanced through the usage of artificial intelligence. This can potentially help companies save time, reduce production costs, mitigate risk factors, and assist in decision-making. Our research indicates a lack of effectiveness in existing agile applications and mentioned numerous approaches to integrate artificial intelligence (AI) into an agile development life cycle which, as discussed, offers better decision-making, automation of routine tasks, and risk analysis that directly impact the overall success of the project.